\documentclass[doublecol]{epl2} 
\usepackage{subcaption}

\title{Supertransient Magnetohydrodynamic Turbulence in Keplerian Shear Flows: The role of the Hall Effect}
\shorttitle{Hal effect in MHD simulations}

\author{Danilo Morales Teixeira\inst{1}\thanks{E-mail: \email{danilo@ita.br}} \and Erico Luiz Rempel\inst{1} }
\shortauthor{Morales Teixeira \& Rempel}

\institute{                    
  \inst{1}Institute of Aeronautical Technology (IEFM/ITA), S\~ao Jos\'e dos Campos - SP 12228-900, Brazil\\
}


\abstract{
Space and astrophysical plasmas are frequently found in the regime of differential rotation, where the presence of a magnetic field can result in the magnetorotational instability, directly responsible for important phenomena such as turbulent angular momentum transport in accretion disks. In the absence of an imposed magnetic field, a nonlinear dynamo is necessary for this transport mechanism to take place. 
In protoplanetary disks there are regions with high density and very low temperatures, which are two necessary conditions for the Hall effect to operate, affecting the development of the dynamo and the associated turbulence. In this work we perform local magnetohydrodynamic (MHD) simulations to study transition to weak turbulence in Keplerian shear flows with Hall effect. The Hall effect is shown to lead the system to long-lived turbulent transients whose decay time follows an exponential dependence on the 
magnetic Reynolds number and the Hall parameter.}

\begin{document}

\maketitle

\section{Introduction}

Transition to turbulence is a widely studied phenomenon in hydrodynamical flows, where usually a control parameter such as the Reynolds number  ($R_e$) is progressively varied and a sequence of local and/or global bifurcations leads the system unto an erratic state where disorder in space and time takes place. In certain situations a subcritical transition is observed, where the laminar state can be shown to be linearly stable for all values of the control parameter and finite amplitude perturbations are required to drive the system away from the laminar flow and unto the turbulent state. This turbulence can be self-sustained or decaying, the former being due to a chaotic attractor in the phase space and the latter due to a {\em chaotic saddle}, i.e., a nonattracting chaotic set \cite{nusse89,rempel07}. 

Several works have studied subcritical transition in hydrodynamical shear-flows like the plane-Couette and pipe flows,  investigating whether the turbulence is sustained or just a long-lived transient. For low values of $R_e$, the flow quickly becomes laminar, whereas for large values it is turbulent. However, for intermediate values of $R_e$ the description of the asymptotic state and the mechanisms for its formation remain an open question \cite{cerbus18}. Recently, the onset of sustained turbulence in shear flow experiments was described in terms of the spatial coupling of transiently chaotic domains, resulting in a second-order phase transition that falls into the directed percolation universality class \cite{avila11,barkley15,lemoult16}.

In astronomy, a typical example of subcritical transition to turbulence is provided by accretion disk models in the absence of magnetic fields and where shear is the only stratification. In such cases, the laminar Keplerian flow is linearly stable \cite{biskamp03}. When a mean magnetic field is present, the magnetorotational instability is a linear instability that can lead the system towards the turbulent state, thus providing a mechanism for outward angular momentum transport and the corresponding inward accretion of matter \cite{balbus91}. Finite magnetic resistive effects can lead to the decay of the magnetic field, so in the absence of an imposed background field a nonlinear dynamo mechanism is required to sustain the magnetic energy and, consequently, the turbulence. Schekochihin et al. \cite{scheko04} conjectured that at low $P_m$ a dynamo might only be possible in the presence of a mean field. In local simulations using the shearing
box formalism with zero net magnetic flux, Fromang et al. \cite{fromang07} reported that the turbulence disappears for a magnetic Prandtl number ($P_m$) below a critical value that seems to be a decreasing function of the kinetic Reynolds number. Iskakov et al. \cite{iskakov07} found that the dynamo action is possible for $P_m<$1 at high Reynolds numbers and/or numerical resolution, but with low growth rates, while Boldyrev \& Cattaneo \cite{boldyrev04} proposed through different assumptions that the dynamo action is possible at any low value of $P_m$ at sufficiently high $R_m$. When the magnetic Prandtl number ($P_m$) is greater than one, it was shown by Rempel et al. \cite{rempel10} that the turbulent state found at intermediate Reynolds number decays with time and the average lifetime of the turbulence grows as an exponential function of magnetic Reynolds number ($R_m$), a behavior named supertransient in Refs \cite{kaneko88,tel08}. Recently, Nauman and Pessah \cite{nauman16} showed that turbulence can be sustained for several thousand shear time units in zero net flux shearing box simulations even when $P_m<1$, provided that extended vertical domains are adopted. Other numerical simulations confirm that the use of taller boxes facilitates the presence of sustained dynamo/turbulence in the shearing box model \cite{shi16,walker17}.

Accretion disks around stars (protoplanetary disks) are weakly ionized because the X-ray radiation of the protostar only ionizes the surface of the disk such that these disks are too cold to provide an effective temperature for thermal ionization in the body of the disk. As a consequence the large regions that are adjacent to the disk mid-plane are not susceptible to the MRI\cite{gammie96} while the MRI is active close to the surface of the disk. These weakly ionized disks are mainly composed of a neutral fluid with a small fraction of the fluid composed by a number of ionized particles of different species. Due to the interactions of these different species, three non-ideal processes are introduced, namely, the Hall effect, ambipolar diffusion and Ohmic dissipation \cite{wardle12}. The Ohmic dissipation dominates in regions with high density and low ionization fraction (inner disk and disk mid-plane), the ambipolar diffusion dominates in the outer region, while the Hall effect is important in between (1--10au) \cite{bai11,simon15}.

The ambipolar diffusion is due to the imperfect coupling of the neutrals with the ionized fluid. Blaes \& Balbus\cite{blaes94} found that when the ion-neutral collision frequency drops below the orbital frequency the MRI is suppressed, a result confirmed by Hawley \& Stone\cite{hawley98}. The Ohmic resistivity is used to model layered accretion disks where their surfaces are sufficiently ionized to allow it to be coupled to the magnetic field and provide conditions for the onset of the MRI turbulence while the mid-plane regions are poorly ionized. Later, the Hall effect can destabilize the plasma depending on the orientation of the magnetic field that otherwise would be stabilized by Ohmic losses\cite{wardle12}.

In this letter we extend the analyses of Rempel et al. \cite{rempel10} through shearing box simulations where the Hall effect has been taken into account to fully characterize whether or not the zero net flux Hall-MHD turbulence is sustained or transient. By computing the mean decay time for different initial conditions as a function of the magnetic Reynolds number and the strength of the Hall effect, we conduct a series of statistical studies at transition to weak turbulence. Our results confirm that the exponential dependence of the transient is observed in Low-$R_e$ simulations, but the Hall effect reinforces the production of magnetic energy, therefore, a sustained dynamo might be expected even for $R_e$ below the critical value found in the Hall-free regime. In order to facilitate the dynamo onset, we adopt the tall box aspect ratio employed by references \cite{nauman16,shi16,walker17}, although this is probably not relevant for astrophysical discs. From now on, we loosely employ the term {\em turbulence}  to describe irregular regimes that depart from the laminar flow.

\section{The Hall MHD equations}

We solve the MHD equations in the shearing box formalism \cite{goldreich65} using the Snoopy pseudo-spectral code \cite{lesur05,lesur07} including the Hall effect in the induction equation \cite{kunz13}. The equations are solved inside a box which is rotating with angular velocity $\Omega(r_0)$, where $r_0$ is a fiducial radius that determines the location of the centre of the box. Shearing sheet boundary conditions are adopted \cite{hawley95}. The shearing-box equations in Cartesian coordinates, with $\phi\rightarrow y$ and $r\rightarrow r_0+x$, are given by:

\begin{eqnarray}
\partial_t\textbf{v}+\textbf{v}\cdot\nabla\textbf{v}=-\frac{1}{\rho}\nabla P+\frac{\nabla\times\textbf{B}\times\textbf{B}}{\mu_0\rho}-2\mathbf{\Omega}\times\textbf{v}+ \nonumber \\
2\Omega Sx\hat\textbf{x}+\nu\nabla^2\textbf{v},
\label{eq1}
\end{eqnarray}

\begin{equation}
\partial_t\textbf{B}=\nabla\times\left(\textbf{v}\times\textbf{B}-\frac{\nabla\times \textbf{B}\times\textbf{B}}{X_{Hall}}\right)+\eta\nabla^2\textbf{B},
\label{eq2}
\end{equation}

\begin{eqnarray}
\nabla\cdot\textbf{B}=0 \nonumber \\
\nabla\cdot\textbf{v}=0,
\label{eq3}
\end{eqnarray}

\noindent where for a Keplerian disk $\Omega=r^{-3/2}$, $S=-r\partial_r\Omega=(3/2)\Omega$, $\nu$ is the constant kinematic viscosity coefficient, $\eta$ is the constant magnetic diffusivity, $\mu_0$ the magnetic permeability, $\rho$ is the gas density, $P$ is the pressure, $\textbf{B}$ is the magnetic field and $X_{Hall}=\sqrt{\rho}/\ell_H$, where $\ell_H$ is the  
Hall lengthscale, which is given by \cite{kunz13}

\begin{equation}
\ell_H=\left(\frac{m_ic^2}{4\pi Z^2e^2n_i}\right)^{1/2}\left(\frac{\rho}{\rho_i}\right)^{1/2},
\label{eq5}
\end{equation}

\noindent where $m_i$ is the ion mass, $c$ is the speed of light, $Z$ the atomic number, $n_i$ the ion density and $\rho_i$ is the ion mass density.
The fluid velocity is decomposed as $\textbf{v}=\textbf{v}_0+\textbf{u}$, where the steady-state solution is given by the shear flow $\textbf{v}_0=-Sx\hat \textbf{y}$ and $\textbf{u}$ is the perturbation field. The kinetic and magnetic Reynolds numbers mentioned below are given by $R_e= 1/\nu$ and $R_m=1/\eta$, respectively. From now on, we set $\Omega=1$, $R_e=70$
and use $R_m$ and $X_{Hall}$ as control parameters. 

\section{Numerical simulations}

For the simulation domain we chose a box as in Riols et al. \cite{riols13}, with sides $(L_x,L_y,L_z)=(0.7,20.0,2.0)$, which was shown to favour the onset of dynamo. Estimating the mean decay time of the turbulence as a function of $R_m$ and $X_{Hall}$ requires a statistical study based on long time series for thousands of different initial conditions, therefore, the numerical resolution employed in this letter is low. Riols et al. \cite{riols13} argued that when the Hall effect is absent for $R_e=70$, simulations with a numerical resolution $(N_x,N_y,N_z)=(24,12,36)$ are well resolved up to $R_m=500$, however we have verified that this is insufficient for simulations with the Hall effect. 

\begin{figure}[!h]
\includegraphics[scale=0.45]{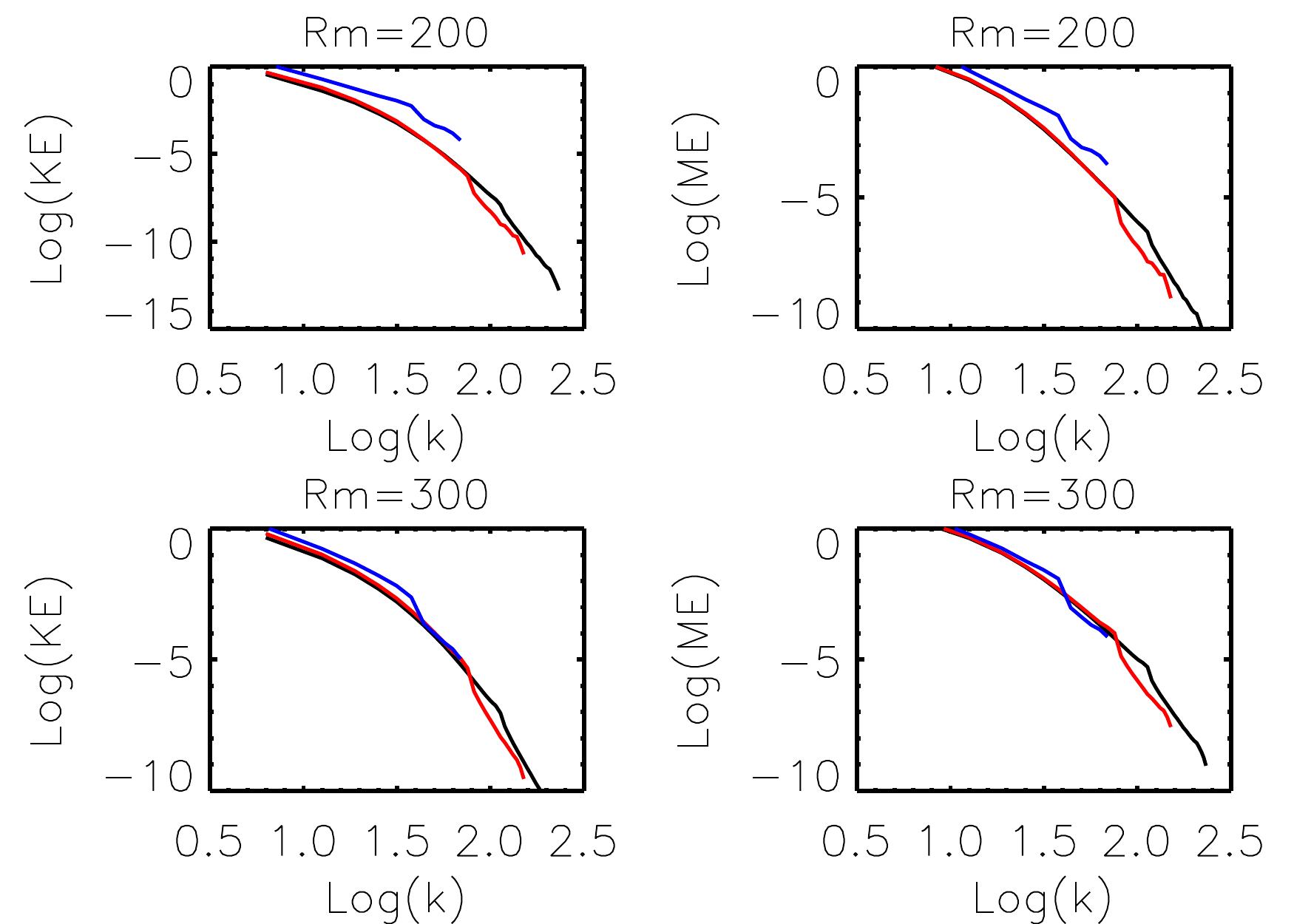}
\caption{Power Spectra of the kinetic (left) and magnetic (right) energies as a function of the wavenumber ($k$) for $R_m=200$ (upper panels) and $R_m=300$ (lower panels). In all case $R_e=70$ and $X_{Hall}=50$. The black lines represent the high resolution simulations, red represents the medium resolution and blue the low resolution.}
\label{fig.1}
\end{figure}

In order to find the minimum resolution required for our study we compute the averages of the Reynolds and Maxwell stresses, as well as the kinetic and magnetic power spectra for $R_e=70$. We test the resolutions $(N_x,N_y,N_z)=(24,12,36)$ (low resolution), $(N_x,N_y,N_z)=(48,24,72)$ (medium resolution) and $(N_x,N_y,N_z)=(72,36,108)$ (high resolution).  In tables \ref{tab.1} and \ref{tab.2} we present the average values of the Reynolds ($<\alpha_{Rey}>=<v_xv_y>$) and Maxwell ($<\alpha_{Max}>=<b_xb_y>$) stresses, averaged over the space and time, for different magnetic Reynolds numbers and in Figure \ref{fig.1} we present the power spectra of the kinetic and magnetic energies for two magnetic Reynolds numbers. Based on our analyses of the average values and the slopes of the power spectra, we conclude that if the medium resolution is adopted, for $X_{Hall}=50$ the values converge up to $R_m=250$. The spectra show some disagreement in low scales, however we have checked that this has low impact on our studies. Similar analysis suggest that for $X_{Hall}=30$ and 40 the values with the medium resolution converge up to $Rm=200$. For this reason we choose the medium resolution for this study and compute the decay times up to $R_m=250$ for $X_{Hall}=50$ and $R_m=200$ for $X_{Hall}=30$ and 40. 

\begin{table}
\caption{Average values of the Reynolds and Maxwell stresses for $R_e=70$, $R_m=200$ and $X_{Hall}=50$.}
\label{tab.1}
\begin{center}
\begin{tabular}{|l|c|r|}
Resolution  & $<\alpha_{Rey}>$ & $<\alpha_{Max}>$\\
\hline
Low & 0.237 & -2.610 \\
Medium & 0.069 & -0.595 \\
High & 0.056 & -0.548
\end{tabular}
\end{center}
\end{table}

\begin{table}
\caption{Average values of the Reynolds and Maxwell stresses for $R_e=70$, $R_m=300$ and $X_{Hall}=50$.}
\label{tab.2}
\begin{center}
\begin{tabular}{|l|c|r|}
Resolution  & $<\alpha_{Rey}>$ & $<\alpha_{Max}>$\\
\hline
Low & 0.182 & -1.287 \\
Medium & 0.103 & -0.798 \\
High & 0.072 & -0.673
\end{tabular}
\end{center}
\end{table}
 
\section{Results}

\begin{figure}[h]
\includegraphics[width=\columnwidth]{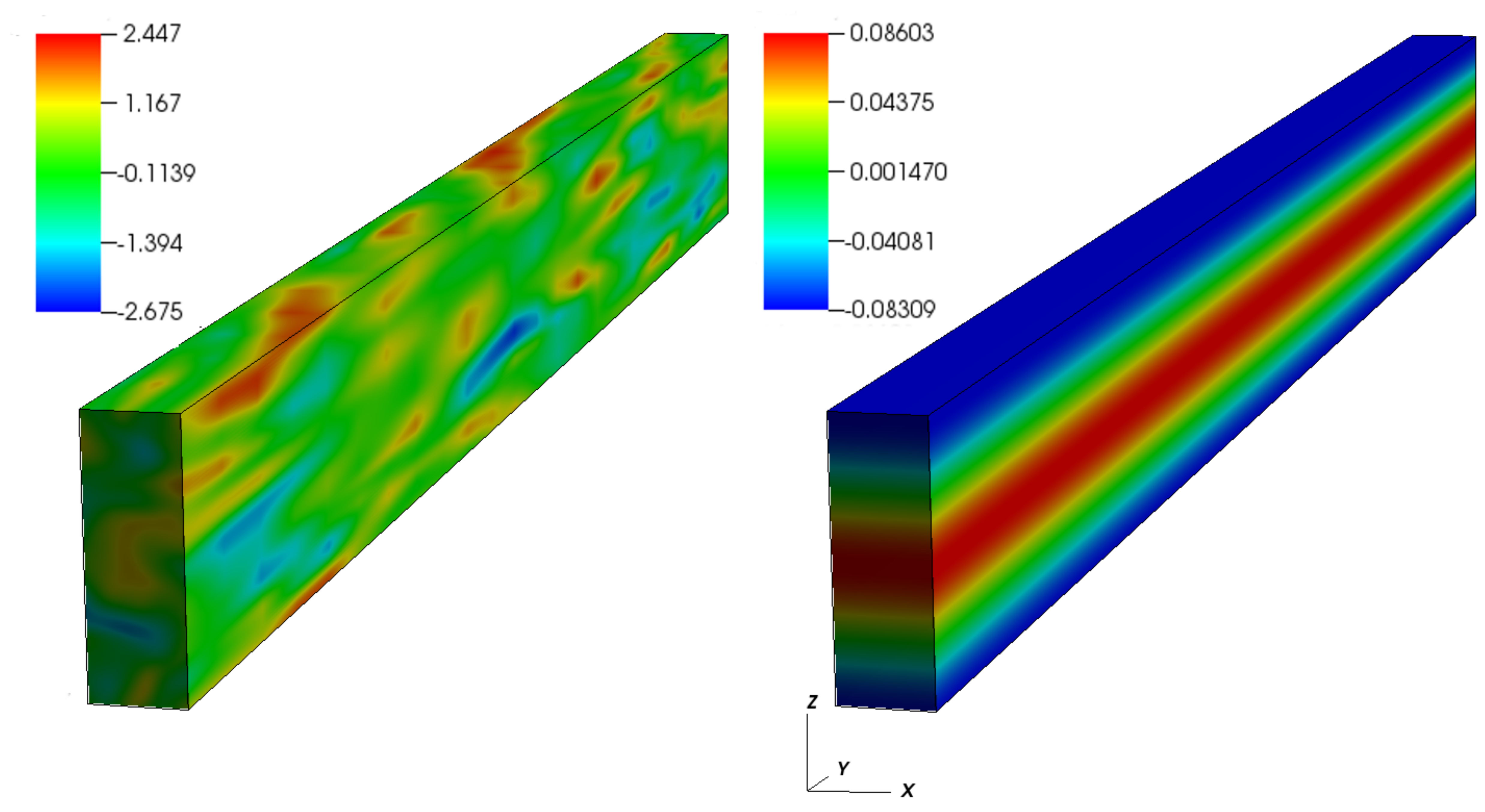}
\caption{Volume visualization of y-component of the magnetic field. Left: Turbulent flow at $t=1000$ shear times. Right: Laminar flow at $t=14000$ shear times.}
\label{fig.2}
\end{figure}

The statistical determination of the mean decay time ($\tau$) requires the integration of Eqs. (\ref{eq1})--(\ref{eq3}) with a large set of initial conditions in turbulent states until they decay to the laminar state. We obtain these initial conditions from long turbulent time series that were generated using $R_e=70$, $R_m=500$ and different values of $X_{Hall}$ between 10 and 1000. The state variables are saved from those time series at each 500 shear time units (before decay) to obtain uncorrelated turbulent initial conditions.

Figure \ref{fig.2} presents snapshots of the azimuthal component of the magnetic field for a simulation with $R_e=70$, $R_m=500$ and $X_{Hall}=50$ at two different times showing a turbulent state in the left panel that decays to the laminar state in the right panel. It is important to emphasize that this simulation is under-resolved and has only been used to generate a long turbulent time series from which we extract different states to use as initial conditions. 

\begin{figure}[!h]
\includegraphics[scale=0.45]{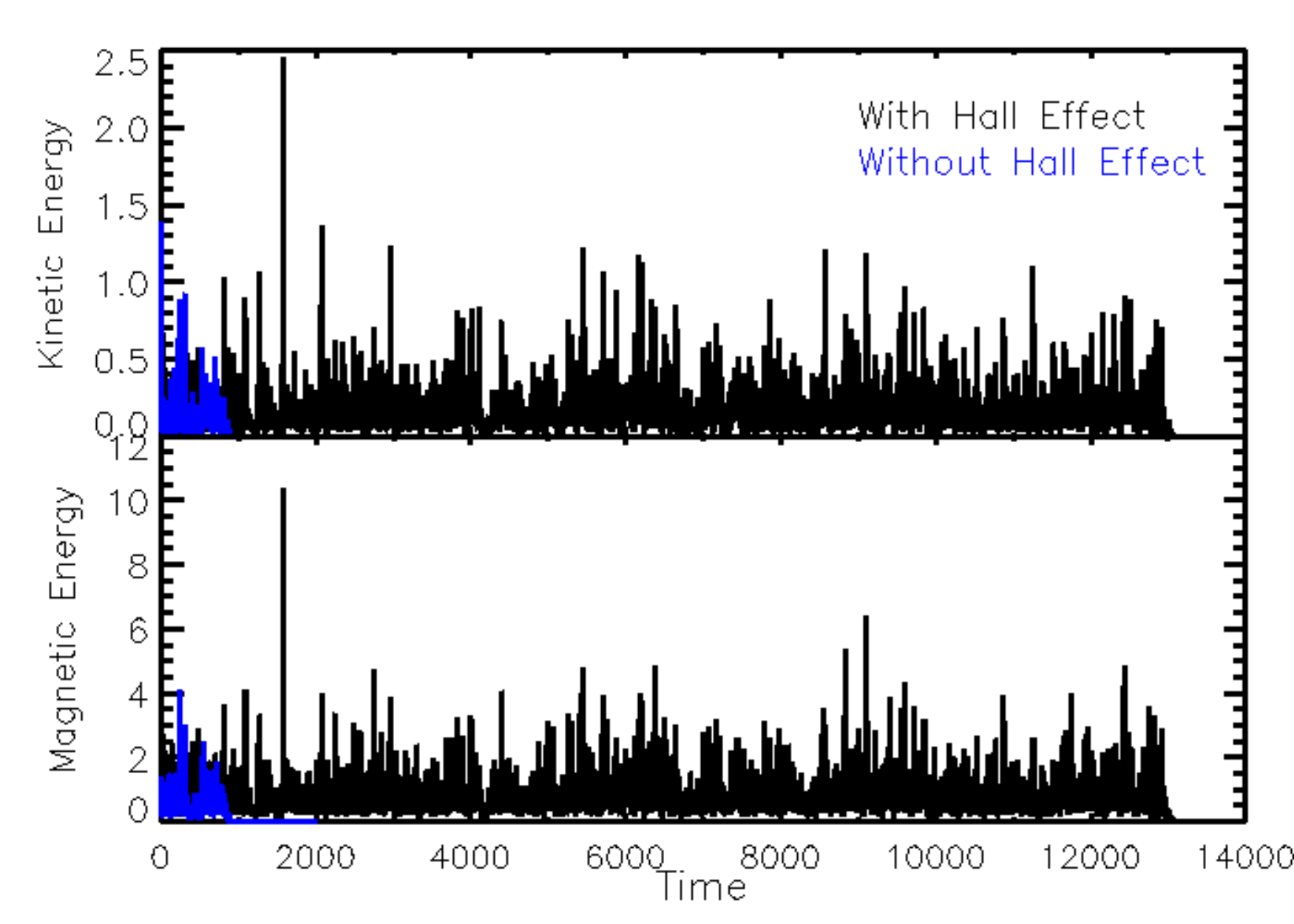}
\caption{Time series of the kinetic and magnetic energies for our medium resolution simulation with $R_e=70$, $R_m=500$ and $X_{Hall}=50$. Top: Kinetic energy as function of the time. Bottom: Magnetic energy as function of the time. The black line corresponds to simulations with Hall effect and the blue line corresponds to simulations without the Hall effect.}
\label{fig.3}
\end{figure}

In figure \ref{fig.3} we show the time series of the kinetic and magnetic energies for two simulations  with medium resolution and $R_e$=70, $R_m=500$ and $X_{Hall}=50$. The black line shows the transition from the turbulent state to the laminar state at $t\approx13.000$ shear time units and corresponds to the Hall effect included in the induction equation; for comparison we are also showing in blue the results from a simulation without the Hall effect, which has transitioned to the laminar state much faster, at $t\approx900$. These results reveal the presence of long turbulent transients whose decay times are increased by the Hall effect. 
%When using higher numerical resolution we observe a corresponding increase in the 
%duration of the turbulent transients, but the system still decays to the laminar state.

\begin{figure}[!h]
\includegraphics[scale=0.45]{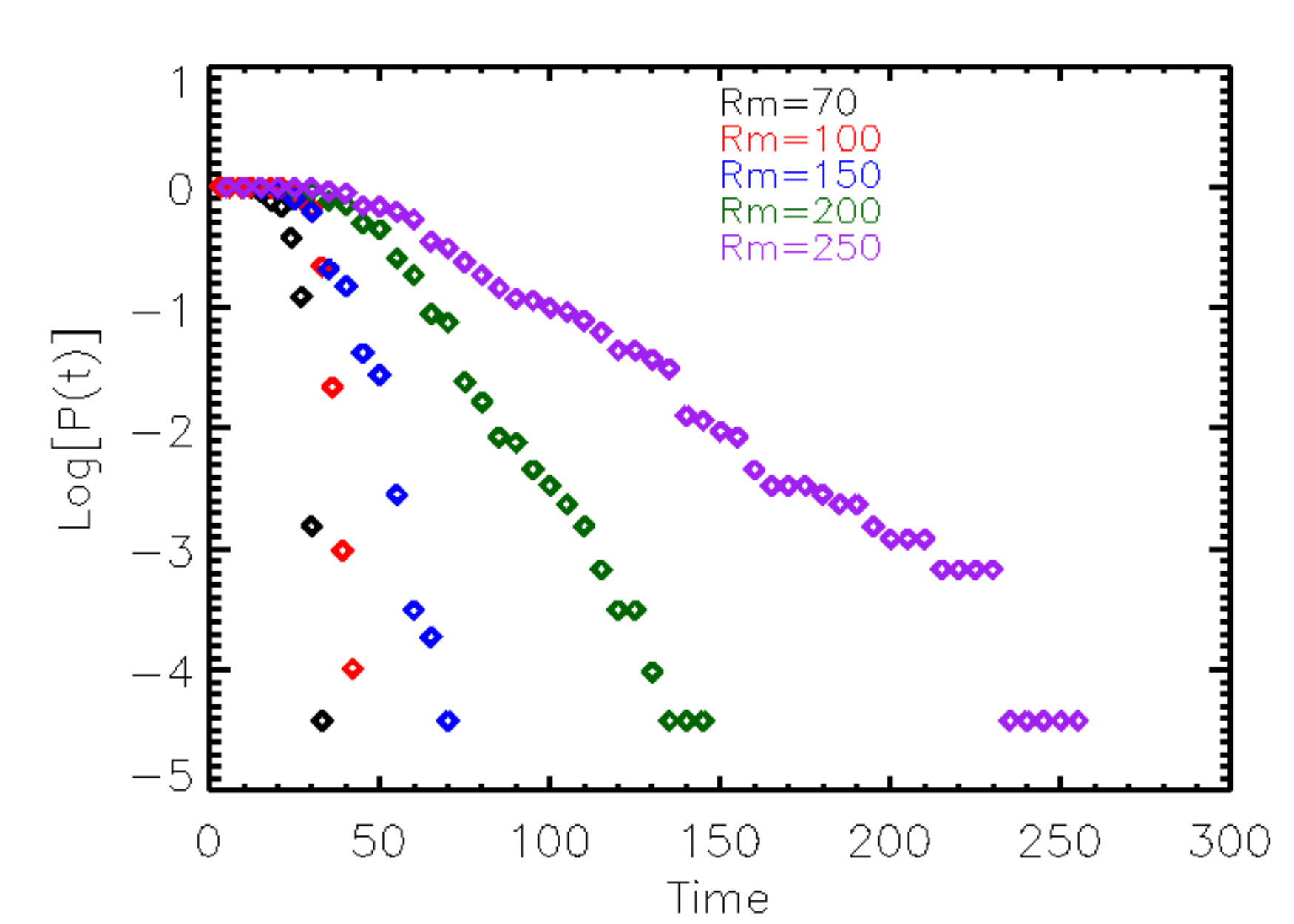}
\caption{Probability ($P(t)$) in log scale of the decay time as a function of the time ($t$) for five different values of $R_m$ and $X_{Hall}$=50.}
\label{fig.4}
\end{figure}

To obtain the average lifetime ($\tau(R_m)$) of the turbulence we calculate the probability ($P(t)$) to find the system in a turbulent state at a corresponding time $t$. For each value of $R_m$ the probability is computed using a set of up to 170 simulations where each simulation corresponds to a different initial condition as described above. The simulations are interrupted when the kinetic energy reaches a level below 10$^{-5}$. Figure \ref{fig.4} shows the probabilities ($P(t)$) in log scale as a function of the time $t$ for different magnetic Reynolds numbers. From these data we perform linear fits to obtain the inverse of the decay time assuming that they have a dependence with the $R_m$ in the form $P(t,R_m)=\exp[-t/\tau(R_m)]$, as expected for transients due to chaotic saddles \cite{hof06}. 

\begin{figure}[!h]
\includegraphics[scale=0.45]{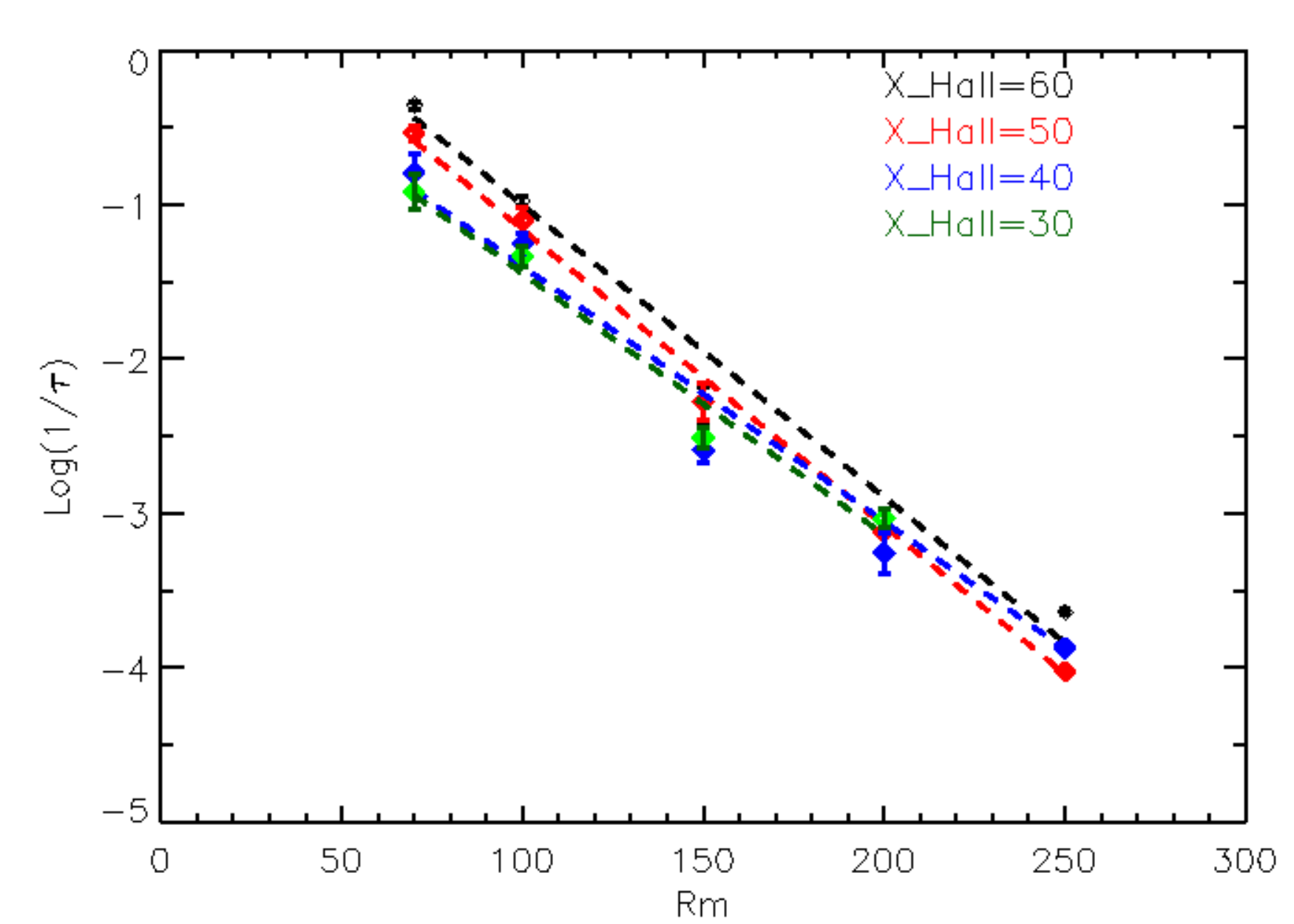}
\caption{Inverse of the decay time for four values of $X_{Hall}$ in log scale as function of $R_m$ with their corresponding linear fits.}
\label{fig.5}
\end{figure}

Using the data obtained from the linear fits we have plotted the inverse of the decay time ($1/\tau$) as a function of $R_m$ and the results are presented in Figure \ref{fig.5}. The inverse of the decay time has a clear exponential dependence with $R_m$. The curves obtained from these linear fits are given by

\begin{equation}
1/\tau=\exp(AR_m+B),
\label{eq6}
\end{equation}

\noindent and the values of the constants $A$ and $B$ for the four $X_{Hall}$ values used in this work are provided in table \ref{tab.3}. Due to the exponential dependence with $R_m$ the corresponding lifetimes of the turbulence follow a supertransient law \cite{kaneko88,tel08}.  We conclude that the Hall effect allows for supertransients to be present even at very low $R_m$, when the system would rapidly decay to the laminar state if the Hall effect were absent. Whether the Hall effect might support the existence of a self sustained dynamo at low $R_m$ numbers or not is the topic of a future work. 

\begin{table}
\caption{Values of the constants $A$ and $B$ in Equation \ref{eq6} for four different values of $X_{Hall}$.}
\label{tab.3}
\begin{center}
\begin{tabular}{|l|c|r|}
$X_{Hall}$  & $A$ & $B$\\
\hline
30.0 & -0.0169 & 0.248 \\
40.0 & -0.0166 & 0.258 \\
50.0 & -0.0192 & 0.760 \\
60.0 & -0.0189 & 0.890
\end{tabular}
\end{center}
\end{table}

\begin{figure}[!h]
\includegraphics[scale=0.45]{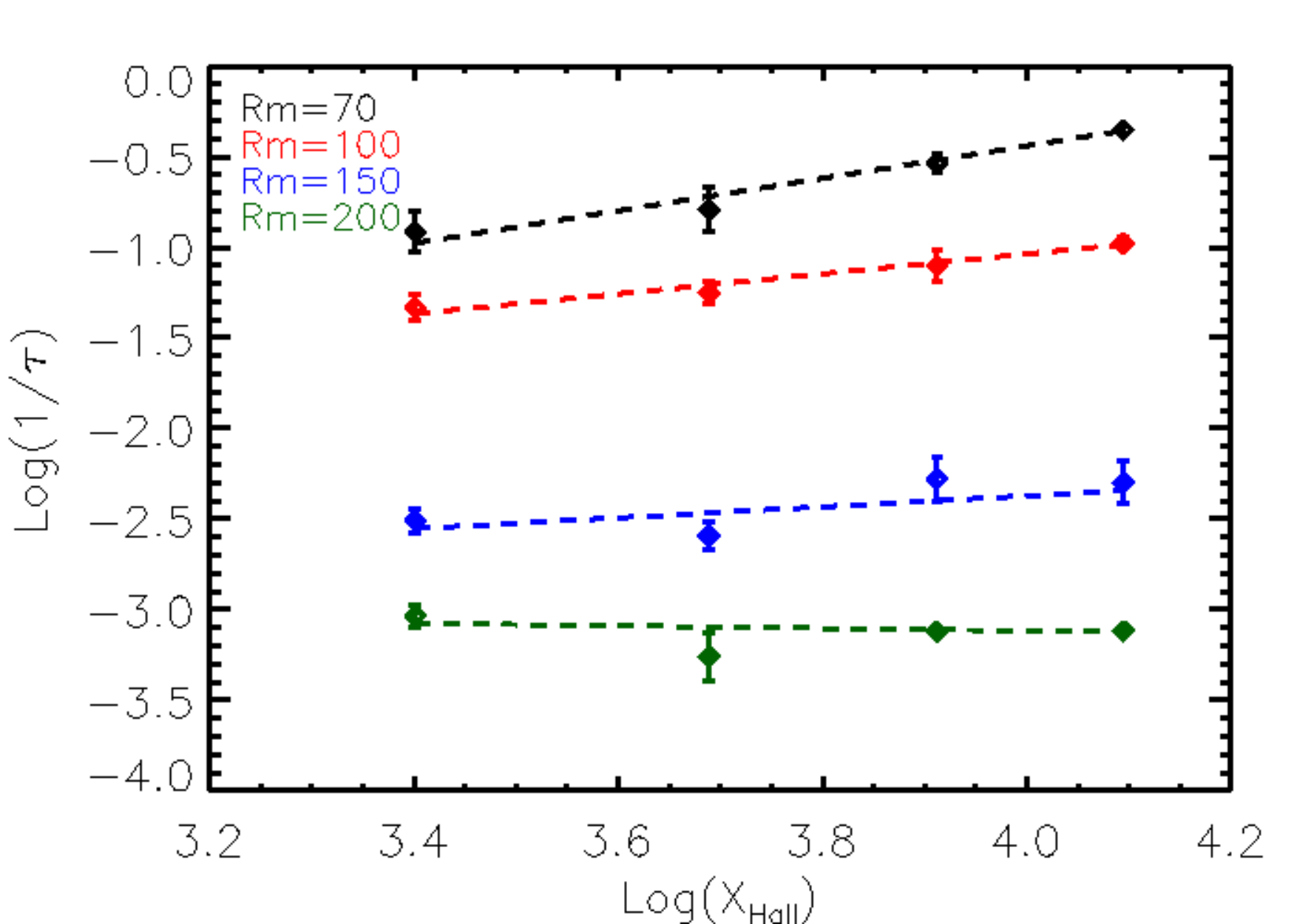}
\caption{Inverse of the decay time for four different values of $R_m$ in log scale as function of $X_{Hall}$ with their corresponding linear fits.}
\label{fig.6}
\end{figure}

The same data from Figure \ref{fig.5} are shown in Figure \ref{fig.6}, but arranged as a plot of the inverse decay time as a function of $X_{Hall}$ for different values of $R_m$. Once again, an exponential law is found from the linear fit in the log--linear plot.

\begin{figure}[!h]
\includegraphics[scale=0.45]{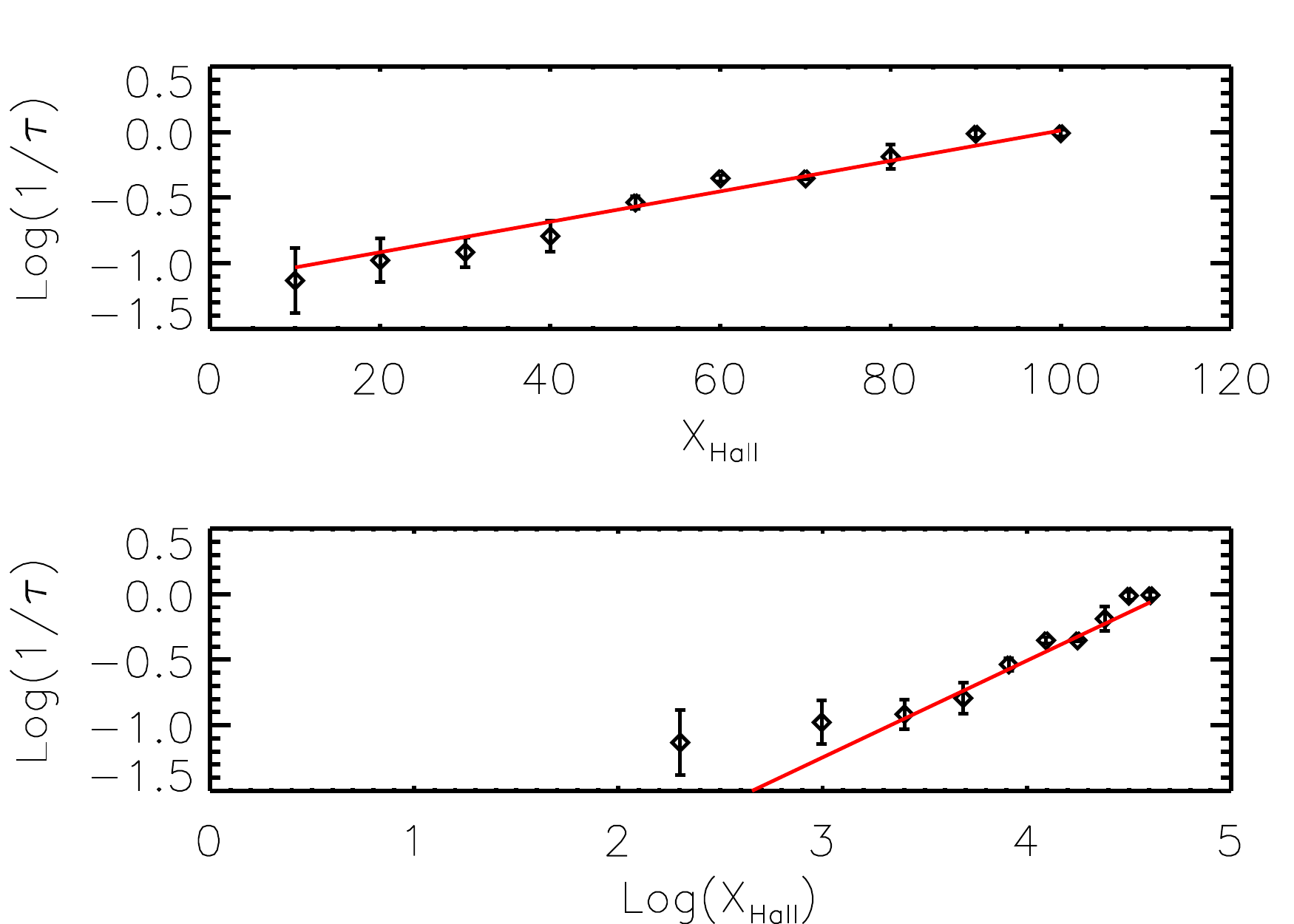}
\caption{Inverse of the decay time for $R_m=70$ as function of $X_{Hall}$. The upper panel presents the values in log-linear scale and the bottom panel the values in log-log scale.} 
\label{fig.7}
\end{figure}

There are two types of supertransient laws described in T\'el and Lai \cite{tel08}, the type-I, given by a power-law, and the type-II, described by an exponential-law. The type-I law is observed from a linear fit in a log--log plot, whereas the type-II comes from a linear fit in a log- linear plot. When someone is using few points in a linear regression it might be difficult to tell which type of supertransient is being observed, since both log-linear and log-log plots seem to allow for a linear fit. To ensure that our data support a type-II supertransient, we show in Figure \ref{fig.7} both the log--linear (upper panel) and log--log (lower panel) plots for $R_m=70$, where it is clear that the log--linear plot provides the best linear fit. Thus, assuming that the inverse of the decay time has a dependence with $X_{Hall}$ in the form

\begin{equation}
 1/\tau=\exp(CX_{Hall}+D),
 \label{eq7}
 \end{equation}
 
\noindent the computed values of the constants $C$ and $D$ are presented in the table \ref{tab.4}. The results for $R_m=200$ may have been affected by inappropriate resolution, thus the slightly negative value of $C$. However, even if one were to disregard it from the analyses, it can be argued that there is a universal type-II supertransient law for the decay time of the MHD turbulence for the parameter space used in this work.

\begin{table}
\caption{Values of the constants $C$ and $D$ in Equation \ref{eq7} for four different values of $R_m$.}
\label{tab.4}
\begin{center}
\begin{tabular}{|l|c|r|}
$R_m$  & $C$ & $D$\\
\hline
70.0 & 0.90 & -4.02\\
100.0 & 0.55 & -3.24 \\
150.0 & 0.30 & -3.58  \\
200.0 & -0.07 & -2.84
\end{tabular}
\end{center}
\end{table}

\section{Conclusion}

Through a series of thousands of long numerical simulations we have characterized the mean decay time of the turbulence  in shearing box Hall-MHD simulations of Keplerian shear flows as a type-II supertransient as a function of both the magnetic Reynolds number and the Hall parameter. The Hall effect is crucial in developing the observed transient turbulence, since in its absence the system quickly decays to the laminar state and this conclusion is supported by the result of Kunz \& Lesur \cite{kunz13}. Nonetheless, we have not found evidence of a self sustained dynamo for the parameters used in our study. Fromang et al. \cite{fromang07} found that the shearing box turbulence does not persist for $P_m<2$ in the Hall-free regime. Our Hall-MHD simulations have explored the range $1\le P_m<4$ and although the Hall-effect is responsible for causing the long transient turbulence, so far even at $P_m>2$ we haven't found evidence of dynamo action. This may be due to the low $R_e$ adopted and the Hall effect itself, but a more detailed study   is required to find out the critical values for $P_m$ for a transition to sustained turbulence. The geometry of the simulation box may also be crucial in allowing for the development of a sustained turbulence as previously mentioned \cite{shi16,nauman16}. A future study with boxes with different aspect ratios is required to 
characterize if this is true for our simulations. Another possibility is that there may be no transition to sustained dynamo and the Hall-MHD turbulence is always transient, with huge decay times due to their exponential growth with the control parameters.

It is worth noting that despite the importance of MRI for accretion processes in astrophysics, certain regions in protoplanetary disks are so poorly ionized that they may not support MRI turbulence \cite{kunz13,riols18}, so that the accretion would take place among a basically laminar state. Nonetheless, recent observations and simulations have revealed the presence of a wealth of complex structures in protoplanetary disks, such as rings \cite{brogan15}, spiral arms \cite{benisty} and vortices \cite{bethune16}. Although the origin of such structures is still a highly debated topic, two non-ideal MHD effects have often been invoked in a tentative explanation for these observations, the Hall effect and ambipolar diffusion. For example, it has been shown in local numerical simulations that in the presence of a vertical background magnetic field the Hall effect can lead to the formation of large-scale self-organized axisymmetric structures called zonal flows \cite{kunz13,riols18}, which correspond to concentric rings in a global disc and are thought to be ideal locations for planet formation \cite{bethune16,riols18}. In Riols et al. \cite{riols18}, zonal flows become more pronounced for stronger background fields; in the absence of a background field, so far we have not observed zonal flows in our simulations.

\acknowledgments
We thank Geoffroy Lesur for his help with the SNOOPY code setup and for helpful discussions. DMT thanks the Brazilian agency CAPES (88887.130860/2016-00) and DMT and ELR thank Brazilian agency FAPESP (2013/26258-4)  for the financial support.

\end{document}